\documentclass[10pt,letterpaper,twocolumn]{article} 

\usepackage[tablesfirst,notablist,nomarkers]{endfloat} 

\usepackage{ol2}
\usepackage[draft]{hyperref}
\usepackage{amsmath}
\begin{document}

\twocolumn[ 

\title{Optical gain in DNA-DCM for lasing in photonic materials}


\author{Marco Leonetti,$^{1,2}$ Riccardo Sapienza,$^2$ Marta Ibisate,$^{2}$ Claudio Conti,$^{3}$ Cefe L\'{o}pez$^{2,*}$ }

\address{
$^1$Dipartimento di Fisica, Universit\`{a} di Roma La Sapienza, \\  I-00185, Roma, Italy
\\
$^2$Instituto de Ciencia de Materiales de Madrid (CSIC) and Unidad Asociada CSIC-UVigo,\\  Cantoblanco 28049 Madrid Espa\~{n}a.2010
\\
$^3$Research Center SOFT INFM-CNR, c/o Universit\`{a} di Roma Sapienza, \\ I-00185, Roma Italy.
\\

$^*$Corresponding author: cefe@icmm.csic.es
}

\begin{abstract}We present a detailed study of the gain length in an active medium obtained by doping of DNA strands with DCM dye molecules. The superior thermal stability of the composite and its low quenching, permits to obtain optical gain coefficient larger than 300 cm$^{-1}$. We also show that such an active material is excellent for integration into photonic nano-structures, to achieve, for example, efficient random lasing emission, and fluorescent photonic crystals.
\end{abstract}


 ] 
Nano-engineered devices and meta-materials are offering nowadays novel ways of controlling light propagation and amplification ranging from  random systems \cite{wiersma} to photonic crystals \cite{Vlasov}.

Nano-structured lasers are typically realized on a sub-wavelength structured dielectric matrix, either periodically or randomly, in which an active medium is inserted to provide optical gain. Their optimization has often being focused on the increase of the matrix quality and refractive index contrast, while recently more efficient active media like fluorescent polymers \cite{D.ZhangFP} have proven to enhance light amplification.

Within these novel materials, DNA strands intercalated with dye molecules  have been proposed as efficient and stable gain medium \cite{A.J.StecklPhoton}.
The superior thermal stability of the composite (up to 250 C$^\circ$) and its low quenching due to a controlled proximity of the dye molecules attached to the DNA provide an enhancement of the emitted integrated luminescence with respect to conventional polymer-Dye composite \cite{A.J.Steckl}. This makes DNA based dyes optimal candidates to infiltrate nanostructures and realize novel active photonic devices.

Direct gain measurements are the best way to compare the optical amplification and efficiency of active media \cite{Costela,Dalnegro}. At variance with simple luminescence experiments, gain studies are not affected by sample thickness uncertainty or by variations of out-coupling due to edge roughness.

In this Letter we  report on the characterization of  the optical amplification of DNA films intercalated with DCM laser dye (DNA-DCM) by measuring the optical gain coefficient $g$ which reaches values as large as 300 cm$^{-1}$. We demonstrate how this material can be used to introduce efficient gain into self-assembled photonic nano-structures.

\begin{figure}[htb]
\centerline{
\includegraphics[width=8.3cm]{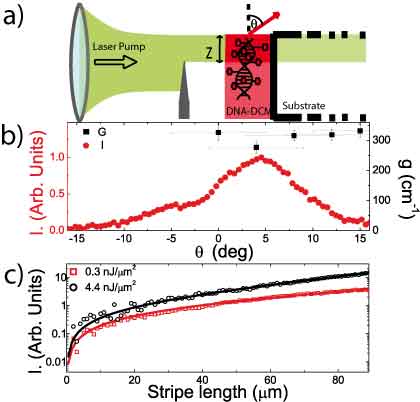}}
\caption{ \label{F1} (Color online). a) VSL setup; b) Intensity from the illuminated  stripe (solid circles) and optical gain coefficient g (solid squares) as function of the detection angle $\theta$; c) VSL measurement at two different pump energies. The continuous lines are fit from the model (see text).}
\end{figure}

For the sample preparation DNA-surfactant insoluble in water is dissolved in ethanol and mixed with dye ethanolic solution \cite{YutakaKawabe}. DNA-DCM films have been obtained by vertical deposition on a glass substrate in an oven at 45 C$^\circ$. Samples with different DNA to DCM weight ratio (ranging from 0.5 $wt\%$ to 10 $wt\%$) and various thickness (10-50 $\mu$m) have been prepared.

The optical gain coefficient has been measured with the  variable stripe length (VSL) technique \cite{Dalnegro} (a sketch of the setup is given in figure 1a). The sample is optically pumped by laser light (0.02--2 mJ/pulse, 9 ns pulse at 10 Hz repetition rate, 532 nm wavelength) focused by a cylindrical lens to a narrow ($\sim$ 20 $\mu$m) stripe. The length $z$ of the illuminated area is adjusted by cutting the pump beam with a blade of variable position. The light emitted  at the edge of the sample, in the direction $z$ of the stripe is focused by a spherical doublet lens (NA = 0.16) to a fiber-coupled spectrograph. The optical gain coefficient $g$ is obtained by fitting the measured emitted intensity $I(z)$ integrated in a wavelength range of 8 nm around the emission maximum at 620 nm, (figure 1c) with :
\begin{equation}
I(z) = \frac{A_0}{g}\, \left[\, \exp(\, g\,z)-1\right].
\label{1}
\end{equation}
In Eq. 1, $g$ is the modal gain defined  as in ref. \cite{Dalnegro} and $A_0$ is a scaling factor accounting for the number of excited molecules and the collection efficiency.

Figure 1b (solid circles) shows the emission intensity as function of the detection angle $\theta$. The asymmetric emission is due to the weak waveguiding effect in the stripe (of refractive index n$\sim$1.53) \cite{Costela}. Solid squares in figure 1b shows that gain does not depend on $\theta$. We performed all measurements by integrating the signal between $\theta=0$ deg and  $\theta=10$ deg.

Figure 2a shows the optical gain coefficient $g$ as function of the DNA-DCM weight percentage. Optimal amplification efficiency is obtained for a dye density between 2 and 3 $wt\%$   while at higher concentrations quenching phenomena  decrease the  effective optical amplification. This value slightly differs from ref. \cite{A.J.Steckl} where maximal light emission is obtained at 1 $wt\%$. It is important to notice that these gain measurements are not affected by uncertainty in the sample thickness or out-coupling at the sample edge which are all included in the prefactor $A_0$.
Figure 2b shows the optical gain coefficient as function of pump power for an optimal DNA-DCM ratio of 3 $wt\%$. For energies higher than  4 nJ/$\mu$m$^2$ the gain coefficient saturates while above 9 nJ/$\mu$m$^2$ gain reduction is due to optical damage. It is remarkable that the maximum measured value for $g$ ($\sim$ 300 cm$^{-1}$), is one order of magnitude greater than that reported for doped plastic materials \cite{Costela}.

\begin{figure}[htb]
\centerline{
\includegraphics[width=8.3cm]{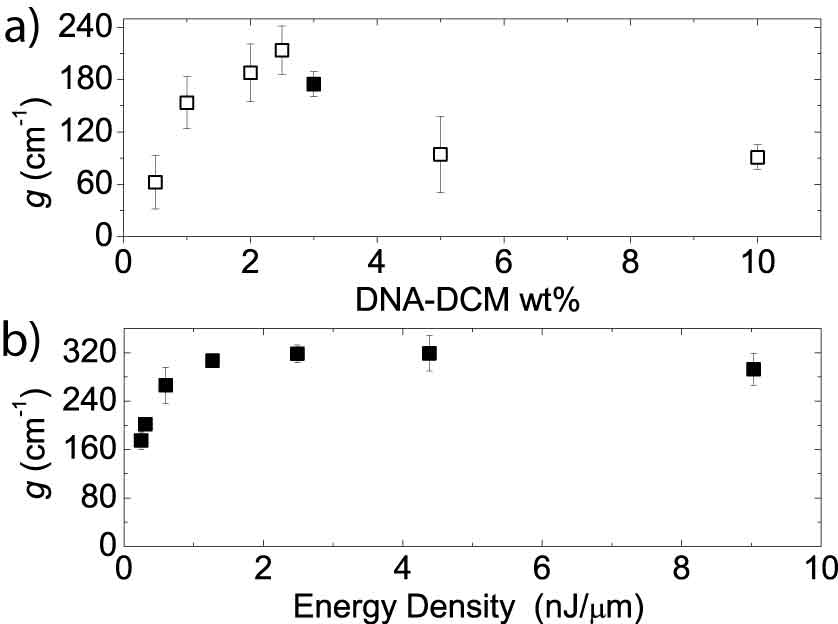}}
\caption{ \label{F2} Optical gain coefficient $g$ plotted as function of DNA-DCM weight percentage (a) and of pump pulse energy density (b) for a DNA to DCM  ratio of 3 $wt\%$ (black dot in fig 2a). }
\end{figure}

The synthetized DNA-DCM is in the form of a solution that can be dried into a film but also easily infiltrated in nano-structured dielectric materials or added at their assembly stage. We followed both strategies. Due to the high polarity of the compound, it destabilizes colloidal solutions of polymeric beads like polystyrene or PMMA. This prevents the formation of a stable colloidal crystal but helps the deposition of active polymeric photonic glasses  \cite{Garcia_PG} made by micron-sized polystyrene beads arranged in disordered fashion. Conversely  it does not affect colloidal silica solutions from which  active photonic crystals  \cite{Cefe_rew} can be efficiently grown. Post-assembly infiltration is feasible but less efficient  due to high viscosity of the DNA-DCM composite. We have grown photonic crystals and photonic glasses with DNA-DCM (dye content of 3 $wt\%$ ). 

Two different structures have been realized by vertical deposition from a water solution containing  450 nm silica beads with and without a 0.2 $wt\%$ of DNA-DCM. 
Figure 3a compares the reflectivity in the $\langle 111 \rangle$ direction for the two samples:  the Bragg peak is shifted by $\sim$30 nm  by the presence of a homogeneous layer of DNA-DCM polymer coating the spheres (confirmed by scanning electron microscopy inspection) that modifies the medium average refractive index. The infiltration process changes the filling fraction but not the
quality of the opal lowering the peak reflectivity to  $\sim$ 0.4 due to contrast decrease, but essentially preserving its linear optical properties.
\begin{figure}[htb]
\centerline{
\includegraphics[width=8.3cm]{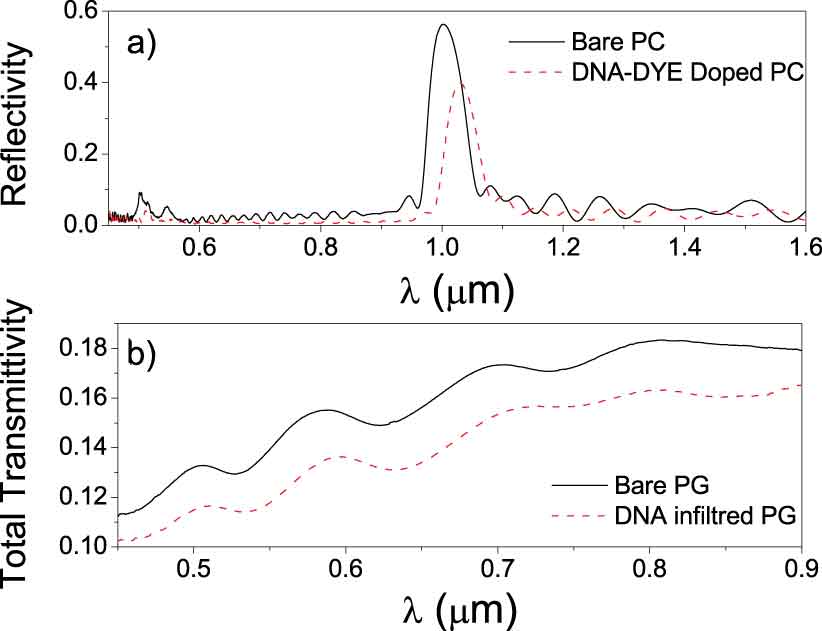}}
\caption{ \label{F3} (Color online). a) Reflectivity as function of wavelength  measured for infiltrated (dashed line) and bare (continuous line) photonic crystal. The Bragg peak is evident around 1 $\mu$m ; b) Total transmissivity  as a function of wavelength in bare (continuous line) and DNA-filled (dashed  line), photonic glass.}
\end{figure}

DNA-DCM can also be efficiently infiltrated in disordered nano-structured system like photonic \cite{Garcia_PG} glasses made of 920 nm polystyrene beads, as the presence of DNA helps the disordered flocculation of the beads.
Total trasmittivity of  the samples with pure DNA (1 $wt\%$) and with only  HCl  as flocculation agent are compared in figure 3b.  This test is performed without DCM to avoid the effect of the large dye absorption.
The characteristic Mie resonances in the total transmission \cite{R.SapienzaPG} are unperturbed by the presence of a small quantity of DNA coating the spheres.
Having proven the feasibility of a non-destructive DNA-DCM infiltration, we have performed a lasing experiment on samples, with and without DNA, obtaining  random lasing (RL) in photonic glasses, but not photonic cystal lasing due to the low refractive index contrast.
\begin{figure}[htb]
\centerline{
   \includegraphics[width=8.3cm]{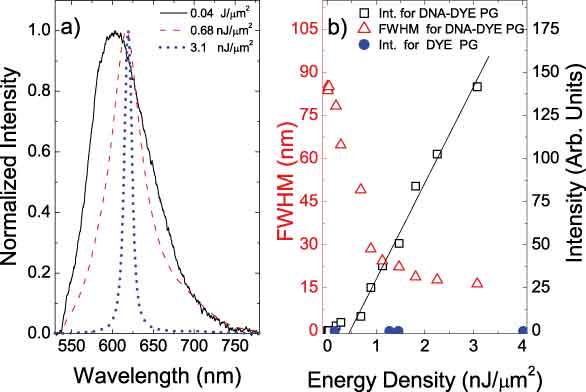}}
      \caption{\label{F4} (Color online). Random lasing from DNA-DCM infiltrated Photonic Glass. a) Emission spectra  from DNA-DCM infiltrated photonic glass at different pump power, below and above the random lasing threshold ($\sim$ 0.5 nJ/$\mu$m$^2$). b) Peak emission (open squares) and FWHM (open triangles) as a function of pump energy for DNA-DCM infiltrated Photonic Glass. Continuous line results from a linear fit to the points above threshold (0.5 nJ/$\mu$m). As a reference, the filled circles plot the peak intensity of a photonic glass infiltrated only with DCM.}
\end{figure}
RL, is the process of emission of stimulated light from a disordered active dielectric medium, which may be produced upon sufficiently intense laser pumping \cite{DiffusiveRLWiersma,Conti}. RL threshold is determined by the scattering mean free path ($\ell_s$) and the gain length ($\ell_g=1/g$); its minimization is achieved by reducing the critical length necessary for net amplification, which is defined in the diffusive approximation as $L_c=\pi \sqrt{\ell_g \ell_{s}/3}$ \cite{Cao_waves_RM}. For the photonic glass infiltrated by DNA-DCM (1 $wt\%$ ),  $\ell_g$ is $\sim $ 40 $\mu$m while $\ell_s$ ranges between 1 and 3 $\mu$m, thus $L_c$ results to be $\sim$ 10-20 $\mu$m. This value for $L_c$, smaller than the thickness of our sample ($\sim$ 50 $\mu$m), allows for RL emission when pumping the DNA-DCM infiltrated photonic glass  with the pump laser focused to a 300 $\mu$m spot. In figure 4a the normalized emission intensity is plotted versus wavelength for different pump pulse energies. The spectral full width at half maximum (FWHM) narrows from 85 nm to 15 nm by increasing the pump energy.  Figure 4b shows the  peak intensity (open squares) and FWHM  (open triangles) as a function of the pump energy. We have obtained a clear efficiency advantage as lasing threshold occurs at 0.5 nJ/$\mu$m$^2$. This is at least one order of magnitude lower than that obtained in previous similar experiments with dry photonic glasses infiltrated with $\sim 10$ times more DCM dye \cite{S.GottardoNatphot}. An identical sample infiltrated with the same amount of DCM (without DNA) does not show significant amplification (fig. 4b filled circles) or line narrowing.

In conclusion, we reported on DNA-DCM infiltrated in nanostructured photonic materials. We measured optical gain coefficients with values as high as $\sim$ 300 cm$^{-1}$, low quenching and a large thermal stability.
Careful characterization of the gain length in the active material allows to predict the lasing threshold and to efficiently design novel light emitting disordered or ordered nano-structures.

The authors wish to thanks R. Feuillade and J. Saenz for the fruitful collaboration. C.C. acknowledges support from ERC GRANT (FP7/2007-2013) n.201766. This work was partially funded by the Spanish MICINN under contract MAT2006-09062 and Consolider NanoLight.es CSD2007–0046. M.I. is a RyC researcher.


\end{document}